\documentstyle[12pt,aas2pp4]{article}
\def\insfig#1{#1}
\def\endinsfig{\end{document}}

\font\smallrm=cmr8

\def\FWHM{{\smallrm FWHM}}

\lefthead{Tonry}
\righthead{Gravitational Lens Redshifts}

\begin{document}

\title{Redshifts of the Gravitational Lenses MG~0414+0534 and MG~0751+2716\altaffilmark{1}}

\author{John L. Tonry}
\affil{Institute for Astronomy, University of Hawaii, Honolulu, HI 96822}
\affil{Electronic mail: jt@avidya.ifa.hawaii.edu}
\authoremail{jt@avidya.ifa.hawaii.edu}

\author{Christopher S. Kochanek}
\affil{Harvard-Smithsonian Center for Astrophysics, Cambridge, MA 02138}
\affil{Electronic mail: kochanek@cfa.harvard.edu}
\authoremail{jt@avidya.ifa.hawaii.edu}

\altaffiltext{1}{Based on observations at the W. M. Keck Observatory,
which is operated jointly by the California Institute of Technology
and the University of California}

\begin{abstract}
We report redshifts in two gravitational lens systems, MG~0414+0534 and 
MG~0751+2716.  The lens galaxy in MG~0414+0534 lies at $z_l=0.9584\pm0.0002$. 
The luminosity and extreme red color of the lens are then typical of a 
passively evolving, early-type, $\sim 2L_*$ galaxy.  The galaxy cannot have a 
significant global mean extinction without being anomalously luminous.  The lens
galaxy in MG~0751+2716 has a redshift of $z_l=0.3502\pm0.0003$ and it is a 
member of a small group.  The group includes the nearby, bright companion 
galaxy whose redshift we confirmed to be $z=0.3501\pm0.0001$ and a nearby 
emission line galaxy with $z=0.3505\pm0.0003$.  A second emission line galaxy
with $z=0.5216\pm0.0001$ was found nearly superposed on the first emission
line  galaxy.  The source in MG~0751+2716 is a $z_s=3.200\pm0.001$ radio quasar. 
For flat universes with $\Omega_0=1.0$ ($0.3$), 96\% (87\%) of lenses like
MG~0414+0534 and 7\% (3\%) of lenses like MG~0751+2716 are expected to have
lower lens redshifts than observed. 
\noindent \it{Subject headings:} cosmology --- distance scale ---
gravitational lensing --- quasars: individual (MG~0414+0435, MG~0751+2716)
\end{abstract}

\section{Introduction}

This is the second paper in a series of observations of the redshifts,
velocity dispersions, and cluster velocity dispersions of gravitational 
lens systems.  While the $\sim 40$ known gravitational lenses are unique 
tools for studying cosmology and galactic structure and evolution, many 
applications are severely limited by the absence of the redshifts of 
either the lens galaxy or the source.  For example, both redshifts are
required to interpret the time delays created by the different ray
paths as a measurement of the Hubble constant (Refsdal 1964), or to
determine the mass-to-light ratios of the lens galaxies and their
evolution with redshift (Keeton, Kochanek \& Falco 1998).  For all
applications it is essential to completely observe and understand as many 
lenses as possible, lest some observational bias or systematic error lead
to an error in the determination of the Hubble constant and other cosmological
parameters, or the evolution rates and properties of galaxies.

MG~0414+0534 was discovered as a four-image radio lens by Hewitt et
al. (1992).  It has a generic image geometry, and in optical images
the three brighter images are connected by a faint arc image of the
quasar host galaxy (Falco, Leh\'ar \& Shapiro 1997).  The optical and
infrared spectrum is that of a very reddened $z_s=2.639$ quasar
(Lawrence et al. 1995).  The lens galaxy is also red, with a color of
R--I$=1.9$ mag (Falco et al. 1997, McLeod et al. 1998).  Previous
attempts to spectroscopically determine the lens redshift have been
inconclusive (see Lawrence, Cohen \& Oke 1995).  A succession of
studies have concluded either that the lens galaxy is very dusty
(Lawrence et al. 1995, Malhotra, Rhoads \& Turner 1997, McLeod et
al. 1998) or that the host galaxy of the quasar is very dusty (Annis
\& Luppino 1993).  The latter interpretation is supported by the
consistency of the photometry of the lens galaxy with a passively
evolving normal galaxy at $z_l \sim 0.8$ (Keeton et al. 1998), the
blue color of the arc (Falco et al. 1997), and the modest differential
extinction between the four quasar images (McLeod et al. 1998).

MG~0751+0435 was discovered by Leh\'ar et al. (1997) as a partial
radio ring lens around the center of a small satellite galaxy named ``G3''
of a larger galaxy G1 located 6~arcsec away and with a redshift of
$z=0.35$.  It was assumed that G3 had the same redshift as G1.
Without a spectrum of the lens itself, the redshift identification the
lens was reasonable but unconfirmed, and the redshift of the source was
unknown.  There are a large number of galaxies near the lens,
suggesting the MG~0751+0435 is another example of a gravitational lens
centered on the member of a small group of galaxies like B~1422+231
(Hogg \& Blandford 1994, Kundic et al. 1997b, Tonry 1998) and
PG~1115+080 (Young 1981, Kundic et al. 1997a, Tonry 1998).

In \S2 we report measurements of the lens redshift in MG~0414+0534 and
both the lens and source redshifts in MG~0751+0435.  We discuss the
consequences of the measurements in \S3.

\section{Observations and Reductions}

MG~0414+0435 and MG~0751+2716 were observed on October 26 and October
27, 1997 using the Low Resolution Imaging Spectrograph (LRIS) (Oke et
al. 1995) at the Keck II telescope on Mauna Kea, as a supplement to
the primary target of double line eclipsing binary stars in M31.
There was light cirrus throughout the run and the seeing was variable,
but extremely good ($<0.75$\arcsec) on the night of Oct 26.  A long
slit of 0.7\arcsec\ width was used for MG~0414+0435 and a slit of
1.0\arcsec\ for MG~0751+2716, along with the 300~l/mm grating blazed
at 5000\AA.  The grating was rotated to provide coverage from
3800--8700\AA with a spectral resolution $\simeq 7.9$\AA\ \FWHM\ and a
scale along the slit of 0.211\arcsec/pixel.  The slit was rotated as
illustrated in Figure 1, at PA 168 for MG~0414+0435 and PA 99 for
MG~0751+2716.  The MG~0414+0435 observation totaled 8000 seconds on
the lens, and then the slit was offset 1.1\arcsec\ east for 2000
seconds to observe the QSO components A and B.  The MG~0751+2716
observation was 4000 seconds and the slit was chosen to cover the
galaxies G1 and G6 as well as the lens (plus QSO), G3.
The observing program was similar in all respects to that of Tonry
(1998), and the same template stars were used for the redshift
determinations reported here.  A log of the observations is presented
in Table 1.

\insfig{
\begin{figure}
\epsscale{1.0}
\plotone{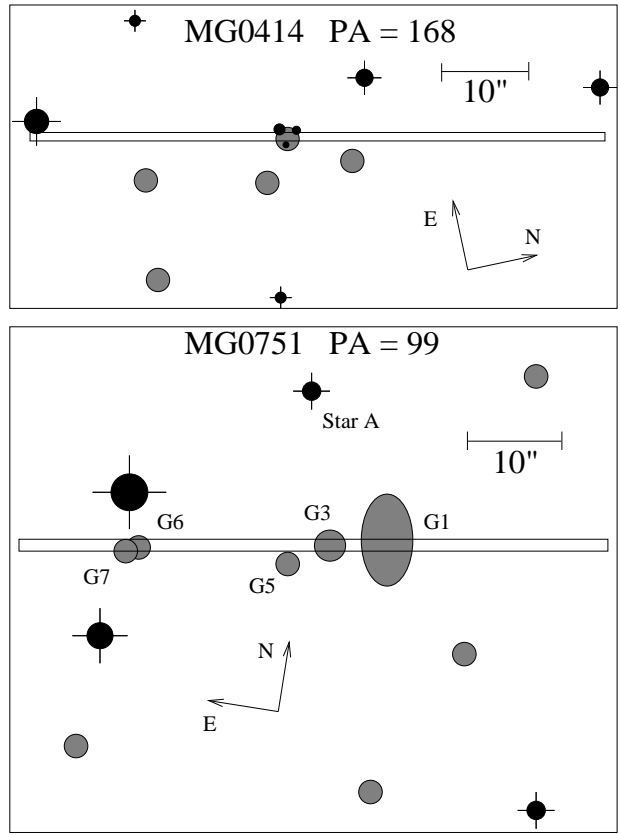}
\caption[fig1.eps]{
Illustration of the slit positions and galaxy identifications for
MG~0414+0435 and MG~0751+2716.  North and East are indicated in each
diagram as well as a 10\arcsec\ scale.
\label{fig1}}
\end{figure}
}

The spectra were reduced using software described in detail by Tonry
(1984).  The basic steps are to flatten the images, remove cosmic
rays, derive a wavelength solution as a function of both row and
column using sky lines (wavelengths tabulated by Osterbrock et
al. 1996), derive a slit position solution as a function of both row
and column using the positions of the template star images in the
slit, rebin the entire image to coordinates of log wavelength and slit
position, add images, and then sky subtract.  A linear fit to
patches of sky on either side of the object (including a patch between
for the galaxy pair observations) did a very good job of removing the
sky lines from the spectra.

The redshift analysis of these galaxies was straightforward since the
spectra were relatively high signal to noise and were essentially 
uncontaminated by
QSO light.  In each case the spectrum was extracted, and
cross-correlated with the template spectrum according to Tonry and
Davis (1979) as well as being analyzed by the Fourier quotient method
of Sargent et al. (1977).  The cross-correlation is more robust in
the case of low signal to noise, but at the signal levels here the two
results are statistically the same, and are simply averaged.  For
each spectrum the redshift and error were calculated.
Table 2 lists the redshifts, errors,
and cross-correlation significance ``$r$'' values for each spectrum or
emission lines used for redshift determination.

\subsection{MG~0414+0435}

As expected from the broad band photometry, the spectrum of the lensing 
galaxy in MG~0414+0435 is very red.  Nonetheless, we clearly detect the 
H and K lines at 7700\AA, which corresponds to a lens redshift of
$z_l=0.9584\pm0.0002$.  The cross correlation $r$ value of 9.1 confirms 
the redshift at very high significance.  The spectrum is illustrated in 
Figure 2.

\insfig{
\begin{figure}
\epsscale{1.0}
\plotone{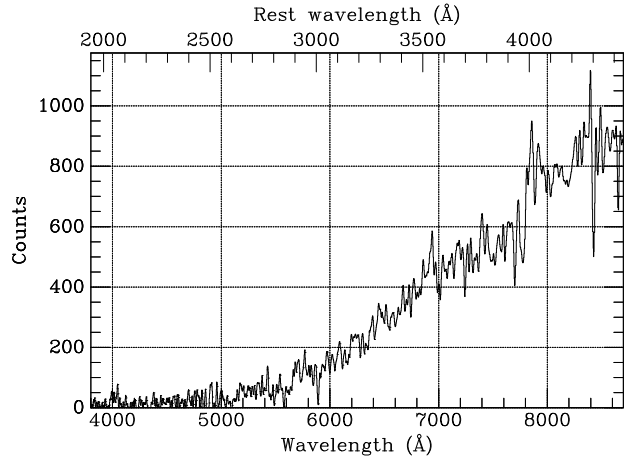}
\caption[fig2.eps]{ The spectrum of the lensing galaxy in MG~0414+0435
is shown.  This spectrum has been Gaussian smoothed with a \FWHM\ of
7\AA.  The top axis shows rest wavelength at a redshift of 0.958.  The
H and K lines appear here prominently at 7750\AA, and the feature at
8400\AA\ is a combination of the G band and a poorly subtracted sky line.
The atmospheric A and B bands have been divided out of the spectrum.
\label{fig2}}
\end{figure}
}

\subsection{MG~0751+2716}

The lensing galaxy, labeled G3 by Leh\'ar et al. (1997), lies approximately 
6\arcsec\ from a much brighter galaxy labeled G1.   Leh\'ar et al. (1997)
measured the redshift of G1 to be $0.351$ and sensibly assumed that the
redshift of the true lens would be the same.  We decided to measure the
redshift of the true lens not only to verify the lens redshift, but
also to attempt to determine the source redshift.  Since the lens is
likely to be the member of a small group, we also chose a slit orientation
which would allow us to measure the redshift of a third galaxy which
we labeled ``G6'' following the notation of Leh\'ar et al. (1997)
(see Figure 1). 

The spectrum which accumulated in 4000 seconds had some surprises.  As
expected, we confirmed the Leh\'ar et al. (1997) redshift for G1 and
found that the lens galaxy shows an extensive absorption line system
at the same redshift.  Superposed on the spectrum of G3, however, was
a strange set of emission lines which turned out to arise from the
QSO.  Figure 3 shows the spectra of G1 and the composite spectrum of the
gravitational lens combined with the source quasar (``GL+QSO'').
The spectrum of G6 also turned out to be composite,
consisting of two galaxies which we label
G6 and G7.  These display prominent 
emission separated by about an arcsecond along the slit at
redshifts of $0.35$ and $0.52$.  The spectra of G6 and G7 are shown in
Figures 3 and 4.

\insfig{
\begin{figure}
\epsscale{1.0}
\plotone{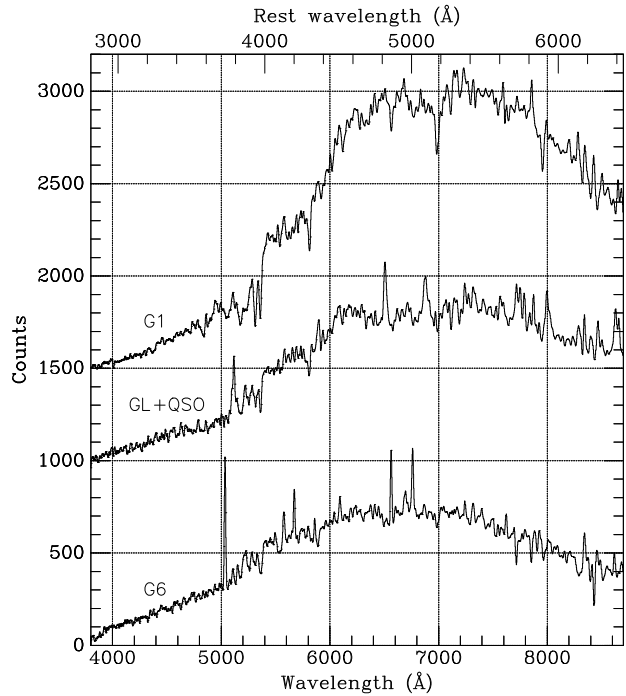}
\caption[fig3.eps]{ The spectra are shown for G1, QSO+lensing
galaxy in the MG~0751+2716 system, and G6.
These spectra have been Gaussian smoothed with a \FWHM\ of 7\AA.  The
top axis shows rest wavelength at a redshift of 0.35.  Emission lines
of 5007, 4959, and 4861 are apparent in the spectrum of G6, as well as
traces of emission from the neighboring G7.  The
strange emission lines in the spectrum of GL, arising from the QSO at
a redshift of 3.2, are shown in greater detail in Figure 4.
For clarity, GL has been offset by 1000 counts and G1 by 1500 counts.
\label{fig3}}
\end{figure}
}

\insfig{
\begin{figure}
\epsscale{1.0}
\plotone{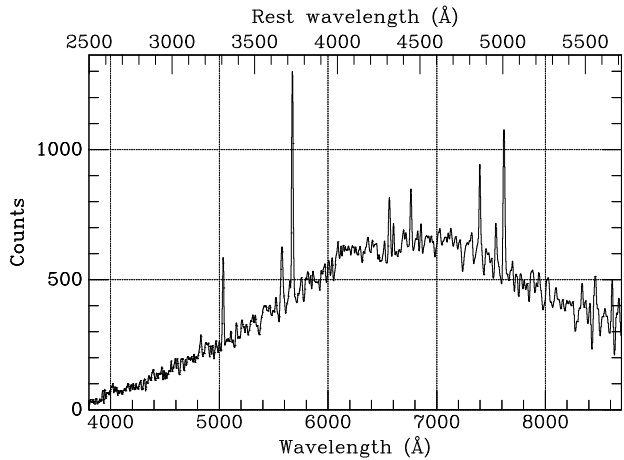}
\caption[fig4.eps]{The spectrum of galaxy G7 (which is superposed on G6)
is shown.  Some traces of the emission from G6 are also present, but
the usual emission of [OII], H, and [OIII] are seen at a redshift of
0.52.
\label{fig4}}
\end{figure}
}

Superposed on the spectrum of 
G3 was a set of five emission lines corresponding to Ly--$\alpha$,
N~IV 1241, C~IV 1549, He~II 1640, and C~III] 1909 at a source redshift 
of $z_s=3.20$.  
Figure 5 shows the spectrum of
the source after subtracting the fraction (0.20) of the spectrum of G1 
which makes the absorption lines of the lens galaxy disappear.  We now
see the emission lines more clearly, as well as Ly-$\alpha$ absorption
on the blue side of the emission line.  
Restricting our attention to 6000-7000\AA\ (R band), we find that this
QSO spectrum has a continuum which is about 0.3 of the total.
Leh\'ar et al. list an R band magnitude for G3 of 21.36 for a
1.4\arcsec\ aperture, so we can estimate an R magnitude of 22.7 for
the QSO and 21.7 for GL itself.

\insfig{
\begin{figure}
\epsscale{1.0}
\plotone{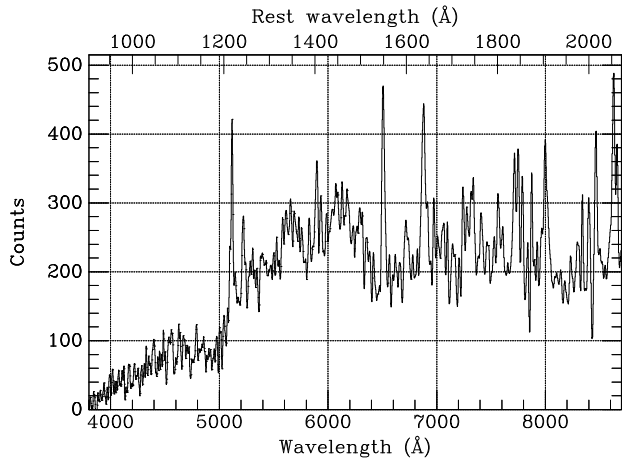}
\caption[fig5.eps]{The spectrum of the QSO, derived as
QSO+GL$-0.2\times$G1, is illustrated here.
These spectra have been Gaussian smoothed with a \FWHM\ of 7\AA.  The
top axis shows rest wavelength at a redshift of 3.20.  Emission lines
of Ly-$\alpha$, C~IV 1549, He~II 1640, and C~III] 1909
are apparent.
\label{fig5}}
\end{figure}
}

\section{Discussion}

There has been a long debate over whether the red colors of the quasar
images in MG~0414 are due to dust in the lens galaxy or dust in
the quasar host galaxy (Hewitt et al. 1992, Annis \& Luppino 1993,
Lawrence et al. 1995, Malhotra et al. 1997, McLeod et al. 1998).  With
the determination of the red color of the lens galaxy (Falco et
al. 1997, Keeton et al. 1998, McLeod et al. 1998) the debate expanded
to include whether its color was simply due to the lens redshift or a
consequence of dust in an otherwise bluer galaxy.  The colors of the
galaxy match those expected for a passively evolving early-type galaxy
at $z_l \simeq 1$ (Keeton et al. 1998, McLeod et al. 1998).  Thus the
spectroscopic redshift of $z_l=0.96$ ends the debate in favor of a
normal, relatively transparent early-type galaxy.  Independent of the
colors of the galaxy, the lens is a $\sim 2L_*$ galaxy in the absence
of any corrections due to internal extinction.  Adding the $A_V \sim
7$ mag of total extinction required to explain the colors of the
lensed images (Lawrence et al. 1995, McLeod et al. 1998) to the lens
galaxy would at a minimum double the intrinsic luminosity to an
implausible $\sim 4 L_*$.
 
In MG~0751+2716 we find another example of a lens galaxy which is a
member of a small group or cluster and where the nearby galaxies make
significant contributions to the lens potential. Unlike the other well
studied systems (Q~0957+561, PG~1115+080, and B~1422+231) where the
lens galaxy is either the brightest galaxy or comparable in luminosity
to the brightest galaxy, the lens galaxy G3 has only $\simeq 13\%$ of
the luminosity of the brightest galaxy G1.  If, however, there is any
surprise in the distribution of lens galaxy luminosities compared to
the other group galaxies, it is probably that systems with larger flux
ratios are not relatively common.  While the low luminosity galaxies
presumably have lower lensing probabilities (for the simple SIS model
the probability is $\propto L^{-1}$) they are also far more numerous
(the luminosity function is roughly $L^{-1} \exp(-L/L_*)$), which
makes the expected distribution of lens galaxy luminosities nearly
flat ($\propto \exp(-L/L_*)$).

The redshifts of lens galaxies relative to their sources can be a
powerful test of the cosmological model (Kochanek 1992), although its
utility as a test is sensitive to the redshift completeness of the
lens sample (Helbig \& Kayser 1996, Kochanek 1996).  For MG~0414+0534
we expect a median lens redshift of $z_l=0.49$ ($0.63$) and a
1-$\sigma$ range of $0.26 < z_l < 0.77$ ($0.36 < z_l < 0.92$) for flat
universes with $\Omega_0=1.0$ ($0.3$).  For MG~0751+2716 we expect a
median lens redshift of $z_l=1.00$ ($1.19$) and a 1-$\sigma$ range of
$0.52 < z_l < 1.52$ ($0.67 < z_l < 1.71$) for flat universes with
$\Omega_0=1.0$ ($0.3$).
\footnote{The 2-$\sigma$ ranges for MG~0414+0534 are $0.11 < z_l <
1.03$ ($0.16 < z_l < 1.19$), and the 2-$\sigma$ ranges for
MG~0751+2716 are $0.21 < z_l < 1.95$ ($0.30 < z_l < 2.18$).}  The lens
redshift of $0.96$ for MG~0414+0534 is high for an $\Omega_0=1.0$
model, wile the lens redshift of $0.35$ for MG~0751+2716 is somewhat
low for an $\Omega_0=0.3$ model.

\acknowledgements
As always, thanks are due to the scientists and engineers responsible 
for the Keck telescope and LRIS spectrograph.  This work was carried
out in collaboration with Tomislav Kundic, and we benefited from his
advice and encouragement.

\clearpage

\begin{deluxetable}{rlrrrrrrr}
\tablecaption{Observing Log.\label{tbl1}}
\tablewidth{0pt}
\tablehead{
\colhead{Obs\#} &  \colhead{Objects} & \colhead{UT Date}  &
\colhead{UT}  & \colhead{$\sec\,z$} & 
\colhead{PA} & \colhead{Exposure} & 
\colhead{Slit} & \colhead{Grating}
} 
\startdata
 1. & MG~0414 GL   & 10/26 & 12:30 & 1.03 & 168 &  2000 & 0.7 & 300/5000 \nl
 2. & MG~0414 GL   & 10/26 & 13:04 & 1.05 & 168 &  2000 & 0.7 & 300/5000 \nl
 3. & MG~0414 GL   & 10/26 & 13:45 & 1.11 & 168 &  2000 & 0.7 & 300/5000 \nl
 4. & MG~0414 GL   & 10/26 & 14:20 & 1.20 & 168 &  2000 & 0.7 & 300/5000 \nl
 5. & MG~0414 A+B  & 10/26 & 14:56 & 1.33 & 168 &  2000 & 0.7 & 300/5000 \nl
 6. & MG~0751 GL   & 10/27 & 13:22 & 1.21 &  99 &  2000 & 1.0 & 300/5000 \nl
 7. & MG~0751 GL   & 10/27 & 14:00 & 1.11 &  99 &  2000 & 1.0 & 300/5000 \nl
 8. & Twilight sky& 10/28 & 04:12 &      &  90 &    20 & 0.7 & 300/5000 \nl
 9. & HD227757    & 10/28 & 04:20 & 1.04 &  90&$5\times5$&0.7& 300/5000 \nl
\enddata
\tablecomments{PA is east from north, exposures are in seconds,
slit widths are in arcseconds.}
\end{deluxetable}

\begin{deluxetable}{lrrrc}
\tablecaption{Redshifts and Emission Lines.\label{tbl2}}
\tablewidth{0pt}
\tablehead{
\colhead{Galaxy} & \colhead{$y$} & \colhead{$z$} & \colhead{$\pm$}  &
\colhead{$r$/emission}
} 
\startdata
MG~0414 GL &       & 0.9584 & 0.0002 & 9.1 \nl
MG~0751 QSO&$0.0$  & 3.200  & 0.001  & {Ly$\alpha$,NIV,CIV,HeII}\nl
MG~0751 GL &$0.0$  & 0.3502 & 0.0003 & 5.4 \nl
MG~0751 G1 &$-5.9$ & 0.3501 & 0.0001 & 15.0 \nl
MG~0751 G6 & 21.2  & 0.3505 & 0.0003 & {[OII],H$\beta$,[OIII]} \nl
MG~0751 G7 & 22.0  & 0.5216 & 0.0001 & {[OII],H$\beta$,[OIII]} \nl
\enddata
\tablecomments{Columns: 
Galaxy name, slit position (\arcsec), redshift and error,
cross-correlation $r$ value or emission lines.}
\end{deluxetable}

\endinsfig

\clearpage

\centerline{\bf FIGURE CAPTIONS}
\bigskip

\figcaption[fig1.eps]{
Illustration of the slit positions and galaxy identifications for
MG~0414+0435 and MG~0751+2716.  North and East are indicated in each
diagram as well as a 10\arcsec\ scale.
\label{fig1}}

\figcaption[fig2.eps]{ The spectrum of the lensing galaxy in MG~0414+0435
is shown.  This spectrum has been Gaussian smoothed with a \FWHM\ of
7\AA.  The top axis shows rest wavelength at a redshift of 0.958.  The
H and K lines appear here prominently at 7750\AA, and the feature at
8400\AA\ is a combination of the G band and a poorly subtracted sky line.
The atmospheric A and B bands have been divided out of the spectrum.
\label{fig2}}

\figcaption[fig3.eps]{ The spectra are shown for G1, QSO+lensing
galaxy in the MG~0751+2716 system, and G6.
These spectra have been Gaussian smoothed with a \FWHM\ of 7\AA.  The
top axis shows rest wavelength at a redshift of 0.35.  Emission lines
of 5007, 4959, and 4861 are apparent in the spectrum of G6, as well as
traces of emission from the neighboring G7.  The
strange emission lines in the spectrum of GL, arising from the QSO at
a redshift of 3.2, are shown in greater detail in Figure 4.
For clarity, GL has been offset by 1000 counts and G1 by 1500 counts.
\label{fig3}}

\figcaption[fig4.eps]{The spectrum of galaxy G7 (which is superposed on G6)
is shown.  Some traces of the emission from G6 are also present, but
the usual emission of [OII], H, and [OIII] are seen at a redshift of
0.52.
\label{fig4}}

\figcaption[fig5.eps]{The spectrum of the QSO, derived as
QSO+GL$-0.2\times$G1, is illustrated here.
These spectra have been Gaussian smoothed with a \FWHM\ of 7\AA.  The
top axis shows rest wavelength at a redshift of 3.20.  Emission lines
of Ly-$\alpha$, C~IV 1549, He~II 1640, and C~III] 1909
are apparent.
\label{fig5}}

\end{document}